\documentclass[traditabstract]{aa}
\usepackage{longtable,lscape,amssymb}
\usepackage{graphicx}
\usepackage{amsmath} 
\usepackage{amssymb}
\usepackage{epsfig}
\usepackage{caption}
\usepackage{float}
\usepackage[caption=false]{subfig}
\usepackage{scrextend}
\usepackage{natbib}
\usepackage{array,booktabs}
\usepackage[colorlinks=true,citecolor=blue]{hyperref}
\newcolumntype{C}{>{$\displaystyle}c<{$}}
\usepackage{txfonts}

\bibpunct{(}{)}{;}{a}{}{,} 

\newcommand{\Msun}{{$M_{\odot}$}}
\newcommand{\Lsun}{{$L_{\odot}$}}
\newcommand{\Rsun}{{$R_{\odot}$}}

\newcommand{\gqc}{2MASS J1549-3539}

\newcommand{\fiveothree}{2MASS J1324-5129}
\newcommand{\fiveofive}{2MASS J1457-3543}
\newcommand{\foreifor}{2MASS J1517-3028}
\newcommand{\threetn}{2MASS J1815-3249}

\begin{document}

\title{Characterization of very wide companion candidates to young stars with planets and disks  \thanks{
Based on observations collected at the European Southern Observatory at Paranal, 
under program 103.C-0200(A), and archive data from 074.C-0037(A) and 082.C-0390(A).}
}

\author{
           F. Z. Majidi\inst{1,2}       
  \and S. Desidera\inst{2}
  \and J. M. Alcal\'a\inst{3} 
  \and A. Frasca\inst{4}
  \and V. D'Orazi\inst{2} 
  \and M. Bonnefoy\inst{5} 
  \and R. Claudi\inst{2}
  \and R. Gratton\inst{2}
  \and D. Mesa\inst{2}   
}


\institute{        
  Dipartimento di Fisica, dell'Universit\'a di Roma Sapienza, P.le A. Moro, 5, Roma I-00185, Italy 
  \and INAF-Osservatorio Astronomico di Padova, vicolo dell'Osservatorio 5, 35122 Padova, Italy  
  \and INAF-Osservatorio Astronomico di Capodimonte, via Moiariello 16, 80131 Napoli, Italy
  \and INAF-Osservatorio Astrofisico di Catania, via S. Sofia, 78, 95123 Catania, Italy
 \and Universit\'e Grenoble Alpes, CNRS, IPAG, 38000 Grenoble, France
}

\date{Received  }

\abstract{Discovering wide companions of stellar systems allows us to constrain the dynamical environment and age of the latter. We studied four probable wide companions of four different stellar systems. The candidates were selected mainly based on their similar kinematic properties to the central star using \textit{Gaia} DR2. The central stars are V4046 Sgr, HIP 74865, HIP 65426, and HIP 73145, and their probable wide companions are 2MASS J18152222-3249329, 2MASS J15174874-3028484, 2MASS J13242119-5129503, and 2MASS J14571503-3543505 respectively. V4046 Sgr is a member of $\beta$-Pictoris Moving Group while the rest of the stellar systems are acknowledged as members of the Scorpius-Centaurus association. The selected stellar systems are particularly interesting because all of them are already known to possess a low-mass companion and/or a spatially resolved disk. Identifying wider companions of these stars can improve their eligibility as benchmarks for understanding the formation channels of  various triple systems, and can help us to determine the orbits of their possibly undiscovered inner, wider companions in case of higher multiplicity. By analyzing the X-Shooter spectra of the wide companion candidates of these stars, we obtained their stellar parameters and determined their ages. We find that 2MASS J15174874-3028484 (0.11 \Msun, 7.4$\pm$0.5 Myr), an already recognized pre-main sequence (PMS) member of Scorpius-Centaurus association, is a highly probable wide companion of HIP 74865. 2MASS J13242119-5129503 (0.04 \Msun, 16$\pm$2.2 Myr) is ruled out as a plausible wide companion of HIP 65426, but confirmed to be a new sub-stellar member of the Scorpius-Centaurus association. 2MASS J14571503-3543505 (0.02 \Msun, 17.75$\pm$4.15 Myr) is a probable sub-stellar member of the same association, but we cannot confirm whether or not it is gravitationally bound to HIP 73145. 2MASS J18152222-3249329 (0.3 \Msun, older than 150 Myr) is determined to be a mildly active main sequence (MS) star, much older than members of $\beta$-Pictoris Moving Group, and unbound to V4046 Sgr despite their similar kinematic features. PMS wide companions such as 2MASS J15174874-3028484 might have formed through cascade fragmentation of their natal molecular core, hinting at high multiplicity in shorter separations which can be confirmed with future observations.     
}


\titlerunning{X-Shooter characterization of four new objects}
\authorrunning{Majidi et al.}
\maketitle

\section{Introduction} 
\label{intro}

The characterization of the wide companions of  stellar systems is of great interest for various reasons in astronomy and astrophysics.  While the discovery of our closest neighbor, Proxima, dates back to more than one century ago \citep[][]{innes15}, soon followed by the discovery of its strong similarity in terms of kinematical features to  $\alpha$ Cen \citep[][]{innes26, luyten1925, alden1928}, only recently
it was demonstrated that Proxima
is bound to the $\alpha$ Cen system \citep[][]{kervella17}. The identification of such wide, physically bound binaries is relevant for our understanding of cloud fragmentation and star formation processes. \citet{elbadry19} show that $100-200 \hspace{0.1cm} (au)$ is the separation range below which binaries are more likely to have formed via fragmentation of individual gravitationally unstable disks rather than through turbulent core fragmentation. Wider binaries, on the other hand, appear to have formed during the dissolution phase of young star clusters as they cannot be explained through the star formation process or by dynamical interactions in the field \citep[][]{kouwenhoven10, mc11}. These wider binaries are expected to exist at a young age and subsequently to be destroyed by dynamical interactions with individual stars and giant molecular clouds \citep[][]{wsw1987, caballero2009, jt10}. There is evidence that such systems exist with young ages (e.g. the AU Mic/AT Mic system with separation 0.23
pc), although the fact that young stars are typically members of loose groups or associations leaves ambiguity
between bound systems and comoving objects \citep[][]{alonso15}. Generally, wide pairs are expected to exist with separations up to $\sim$1 pc \citep{cg2004, quinn2009, tian19}.

The study of wide binaries additionally allows us to put constraints on the dynamical environment of a star, which impacts the evolution of the
disk and planetary system around it. On a long (Gyr) timescale, even a distant, low-mass companion (hundreds of
astronomical units) could shape a planetary system through the Kozai mechanism, increasing the eccentricity of the orbit of
the existing planets \citep[][]{kaib13} and possibly producing hot-Jupiters \citep[][]{wang17}. The case of substellar companions at
wide separation around stars that host planets or brown dwarfs at separations close enough to be formed in the circumstellar disk is of great interest. A recently discovered example is the $\epsilon$ Ind system, with a
Jupiter-like planet in a slightly eccentric orbit \citep{feng19}
and a close pair of brown dwarfs at very wide separation \citep{scholz2003,king10}. Such relatively old systems provide us with benchmark cases for studying the formation of gas giant planets and brown dwarfs.

Physical companions can also be used to improve the determination of system age \citep{ghez1993,duchene2007,connelley2008,tobin16},
especially when the central stars bear limited sensitivity to age indicators - as may be the case for both young and old stars, early and late type \citep[e.g., see][]{gcs11, cr12}.
Depending on the mass and evolutionary stage, companions can provide an opportunity to derive the ages of systems thanks to the determination of lithium in their spectra, gravity-sensitive spectral features (such as \ion{Na}{i}, \ion{K}{i}, \ion{Rb}{i}), magnetic activity,
and placement on the color-magnitude diagram (CMD). If the candidate is only a member of the stellar association and not physically bound to its associated star, there is still valuable information to extract from its physical parameters. Identifying new members of an association contributes to a better determination of age and chemical composition of the association. The discovery of multiple systems in young associations is of great value because it offers the opportunity to study objects of different 
mass whose kinematics still reflects the initial conditions of their parental cloud and is likely unaffected by interactions with nearby bodies.

The synergy between \textit{Gaia} DR2 \citep[][]{gaia} and the Spectro-Polarimetric High-contrast Exoplanet REsearch (SPHERE) instrument  represents an important step forward
for the identification of wide companions. \textit{Gaia} has started to provide a wealth of information for the comprehensive identification of bound stellar companions
\citep[][]{andrews17}. The recent DR2 release allows us to further boost this kind
of search, thanks to the extension to much fainter magnitudes, the increased accuracy of parallaxes and proper
motion, and the availability of high-quality photometry and color information and even radial velocities
for a subsample of targets. The sensitivity of \textit{Gaia} is high enough to allow the identification of brown dwarf
companions at wide separations from young stars in the closest star-forming regions and young associations (age
$<$ 20 Myr, dist$\leq$ 150 pc). In association, SPHERE, an extreme-AO instrument installed at the VLT since 2014 \citep[][]{beuzit19}, is providing fascinating results on the
detection and characterization of planetary mass companions, with some cases even being observed in the process of formation \citep[][]{keppler18}. This instrument is also providing an unprecedented view of circumstellar disks \citep[][]{garufi17}
thanks to a variety of instrument modes and its superb image quality as well as improved spatial resolution.
Most of the targets observed with SPHERE are young stars (Desidera et al., A\&A, submitted), because of the increased brightness of substellar objects
at young ages and typical lifetimes of circumstellar disks.

In this context, we performed a pilot program, identifying wide companion candidates around SPHERE-GTO targets with directly imaged substellar companions or spatially resolved disks.
We selected five candidates for this program and the current paper presents the spectroscopic characterization of four of them. An early result of this program, the identification and characterization of a new probable wide companion of GQ Lup was presented in \citet{alcala20}. Here we present additional physical parameters of this target as a complement to the addressed letter and characterize the other four candidates by means of intermediate-resolution VLT/X-Shooter spectroscopy. The outline of the paper is as follows. In Sect. 2, the process of target selection exploiting SPHERE-GTO and \textit{Gaia} DR2 capabilities is explained. In Sect. 3, we describe the spectroscopic observations, data reduction, and the methods of analysis used for the determination of stellar parameters. We present the results of our study in Sect. 4 and discuss the configuration of these systems in the framework of current formation theories in Sect. 5.

\section{Target selection} 
\label{obs}
 
The five wide companion candidates were selected based on \textit{Gaia} DR2 and SPHERE data. Possible wide companions to SPHERE-GTO targets were searched for in the
\textit{Gaia} DR2 catalog, and several tens of candidates were found based on their parallaxes and proper motions. Follow-up spectroscopic observations are therefore mandatory in order to confirm the
young age of the proposed targets (high levels of magnetic activity, low-gravity spectral features), study
the occurrence of accretion, and determine the effective temperature. While we expect at least some of our candidates to be gravitationally bound to their central stars, they might also be physically unbound to the selected system. In the latter case, our targets are either members of the association they are selected from and are incidentally co-moving with the central star, or, in the worst scenario, they are field objects appearing in the association by chance. However, in order to confirm two objects as a physical pair we need more accurate measurements of the radial velocities of both the central stars and their probable wide companions. Also, for some of the targets, further observations of the close, directly imaged companions will allow us to take into account their impact on the kinematic properties of the central stars. Below, we provide a brief description of the individual targets and their specific features. For a summary of the physical properties  of the  central stars see Table \ref{central}, and for the  kinematic properties of the  objects see Table \ref{table:kinematic}. To the best of our knowledge, the central stars selected in this work and their wide companion candidates are not listed in the previously investigated wide binaries in open clusters \citep[see, e.g.,][]{kh2007, dk20}.

For convenience, we introduce abbreviations for the names of our targets: 2MASS J18152222-3249329 is referred to as \threetn \hspace{0.02cm}, 2MASS J15174874-3028484 as \foreifor \hspace{0.02cm}, 2MASS J13242119-5129503 as \fiveothree \hspace{0.02cm}, 2MASS J14571503-3543505 as \fiveofive \hspace{0.02cm}, and 2MASS J15491331-3539118 as \gqc.

\begin{table*}
        \centering
        \caption{Physical parameters of the central stars.}
        \begin{minipage}{\textwidth}
        \begin{tabular}{lccccccccr} 
                \midrule
                Name & Distance \footnote{Distances are calculated based on the parallaxes of the objects reported in \textit{Gaia} DR2 catalog.} & Association & SpT & $T_{\rm eff}$ & $A_v$ & Wide companion & Separation & Separation  \\
                & (pc) & & & (K) & (mag) &  & (\arcsec) & (au) \\
                \midrule
                
                 V4046Sgr & 72.4 & $\beta$-Pictoris MG & K5/K7 & 4370/4100 & 0 \footnote{\citep[][]{stemp2004}} & \threetn \hspace{0.02cm} & 901 & 65232 \\
                 HIP 74865 & 123.53 & Sco-Cen (UCL) & F3V & 6720 & 0 & \foreifor \hspace{0.02cm} & 90 &  11118 \\                
                 HIP 65426 & 109.21 & Sco-Cen (LCC) & A2V & 8840 & 0 & \fiveothree \hspace{0.02cm} & 142 & 15508\\
                 HIP 73145 & 133.65 & Sco-Cen (UCL) & A2IV & 8840 & 0 & \fiveofive \hspace{0.02cm} & 280 &  37422 \\               
                 GQ Lup & 151.82 & Sco-Cen (Lupus I) & K7V & 4070 & 0.7 \footnote{\citep[][]{alcala17}} & \gqc \hspace{0.02cm} & 16 & 2429  \\                  \midrule
        \end{tabular}
        \label{central}
\end{minipage}
\end{table*}

\textbf{\threetn} is a possible additional member of the V4046 Sgr system. The central object is a close binary composed of two young stars orbiting with a period of 2.4 days and still accreting matter from a gas-rich circumbinary disk \citep[][]{stemp2004, rosenfeld13,vale19}. It is also the oldest Classical T Tauri star to date with a spectral type (SpT) earlier than M5  \citep[the
expected age of the system is 24 Myr according to isochrones and its membership to the $\beta$-Pictoris Moving Group; ][]{torres2008}. The wide
companion GSC 7396-0759, instead, hosts a recently discovered debris disk \citep[][]{sissa18}
and is not accreting. The discovery of another component of the system of slightly lower mass than GSC 7396-0759 would provide additional information on disk dispersal timescales. The new candidate also has an X-ray counterpart from XMM \citep[source 3XMM J181522.2-324932 from][]{rosen16}, suggesting
it is an active star. Information on the separation of the object from the central star and its magnitudes is included in Tables \ref{central} and \ref{table:colors}, respectively.  

\textbf{\foreifor} is a possible wide companion of HIP 74865. It is a member of the Sco-Cen
association, has a brown dwarf companion at small separation (0.13\arcsec = 16 au), and was discovered with Sparse
Aperture Masking observations \citep[][]{hinkley15} and was recently characterized with SPHERE
(Cantalloube et al. in prep.). 
The expected mass of the new candidate is about 0.17 $M_{\odot}$ if coeval to the central star.

\textbf{\fiveothree} is a possible, very wide companion of HIP 65426, and hosts a planetary-mass companion recently discovered with SPHERE at about 90 au \citep[][]{chauvin17}. The
system is part of the Sco-Cen association and its expected mass is about 25 $M_{jup}$ if the target is coeval to HIP 65426.

\textbf{\fiveofive} is a possible comoving object to HIP 73145. HIP 73145 is an A2IV star with an age of $\sim$ 15 Myr. This star was confirmed to be a member of the Scorpius-Centaurus association \citep[][]{rizzuto11} and a debris disk with concentric rings around it was  recently spatially resolved by \citet{feldt17}. The large projected separation (280\arcsec, 38000 au) and the kinematic properties of the wide companion candidate compared to those of the central star lower the chances of them being physically associated. However, the excess noise of 0.993 mas with a significance of 3.29 reported for \fiveofive \hspace{0.02cm} in \textit{Gaia} DR2 may
imply some bias in the astrometric parameters. This target is the faintest object in our sample (\textit{G}=18.52, \textit{Ks}=13.11) and has very
red colors.

\textbf{\gqc} is a highly probable wide companion of GQ Lupi, and is a very young star (age $\sim$ 2 Myr)
in the Lupus cloud according to \citet{alcala20}. GQ Lupi hosts a brown dwarf (BD) companion (GQ Lup B; mass about 20-40 $M_{jup}$) at 0.7\arcsec and has a
disk spatially resolved by ALMA \citep[][]{mcgregor17}. The BD companion is a strong accretor, indicating the presence of a disk around it \citep[][]{zhou14}. The full characterization of \gqc \hspace{0.02cm} was carried out by \citet{alcala20} and their results are presented again here for comparison with other candidates. For the sake of completeness, we report the equivalent widths (EWs) of additional lines that are absent from this latter paper. 
\citet{lazzoni20} spatially resolved the disk around the star exploiting HST archive images.

\section{Observations and methods}

\subsection{Observations}

X-shooter spectra provide us with a wide range of wavelengths, which enables us to perform spectral classification and accretion evaluation as suitable indicators are spread in different spectral regions. For details of the instrument description, the reader is referred to \citet{vernet11}. X-Shooter also enables us to determine the radial velocity of objects within a precision range of 2-5\,km/s \citep[][]{frasca17} and is therefore useful
for confirming or ruling out the physical association of the candidates with their parent stars.
Four of our targets were observed through the $1\farcs 0$, $0\farcs 9$, and $0\farcs 9$ slits for one or two cycles based on their \textit{G} and \textit{J} band magnitudes. Among these four, for the three fainter objects we performed two cycles of ABBA nodding mode. These three targets are \fiveothree, \fiveofive, and \gqc \hspace{0.02cm} with \textit{G} magnitudes 18.71, 18.52, and 18.37 respectively. As measuring line fluxes with a precision better than 20\% --at least in the visible (VIS) and near-infrared (NIR) arms-- was a necessity for achieving our aims, a total on-source execution time of 1.2 hr was requested for the aforementioned objects. For \foreifor \hspace{0.02cm}, which is brighter than the other three targets, one cycle of ABBA  nodding mode  was performed with a total execution time equal to 1 hr. The corresponding \textit{G} magnitude for this latter candidate is 15.80. For a summary of the color magnitudes of our targets see Table \ref{table:colors}. 

\threetn \hspace{0.02cm} is the brightest among our five targets with a \textit{G} magnitude of 13.60. We therefore decided to observe it with the highest resolution offered by X-shooter, adopting the slits $0\farcs 5$, $0\farcs 4$, and $0\farcs 4$, amounting to an execution time of 0.7 hr. For all five targets, short exposures (of $\sim$ 10\% of the time allocated to the nodding mode) were performed before science observations through $5\farcs 0$ slits in order to obtain a more accurate flux calibration and to compensate the slit losses; the slit-loss correction factor (SLCF) for each arm is reported in Table \ref{obs}. These short exposures in stare mode were incorporated in the same observation block (OB) consisting of the nodding mode in order to minimize the overheads. For each object, telluric standards were observed exploiting the same nodding strategy as for the targets in order to minimize noise and cosmetics, with an airmass as close as possible to the targets' airmass for telluric correction purposes. The reported airmass and seeing in Table \ref{obs} are averaged over the observation periods for each arm.  All
targets were observed at parallactic angle.  

\begingroup
\setlength{\tabcolsep}{2pt} 
\renewcommand{\arraystretch}{1.5} 
\begin{table*}
        \centering
        \caption{Observation policies information table. Nodding slits, the allocated exposure time to each arm, seeing, and SLCF are reported in order for UVB, VIS, and NIR wavelengths. The number of single observations is presented as a multiplication factor. The column allocated to $T_{tot}$ presents the total execution time for the 
targets.}
\begin{scriptsize}
        \begin{tabular}{lcccccccccr} 
                \midrule
                Name & Date & $\alpha$ (J2000) & $\delta$ (J2000) & \textit{G} & Nodding slits & Exposure time & Seeing & $T_{tot}$ & SLCF & airmass\\
                & (yyyy-mm-dd) & (h:m:s) & (d:m:s) & (mag) & (\arcsec) & (sec) & (\arcsec) & (hour) & &\\
                \midrule
                
                \threetn \hspace{0.02cm} & 2019-05-24 & 18 15 22.23 & $-$32 49 33.07 & 13.6 & $0\farcs 5$,$0\farcs 4$,$0\farcs 4$ & 2$\times$/450/400/450& 0.955/0.875/0.805 & 0.7 & 1.35/1.35/1.35&1.01\\
                \foreifor \hspace{0.02cm} & 2019-05-24 & 15 17 48.75 & $-$30 28 48.42 & 15.8 & $1\farcs 0$,$0\farcs 9$,$0\farcs 9$ & 2$\times$/800/750/800 &1.01/1.01/1.01 & 1 & 1.5/1.5/1.6&1.09\\                
                \fiveothree \hspace{0.02cm} & 2019-07-04 & 13 24 21.18 & $-$51 29 50.39 & 18.71 & $1\farcs 0$,$0\farcs 9$,$0\farcs 9$ & 4$\times$/800/750/800 & 0.905/0.905/0.895 &1.5 & 1.47/1.34/1.45&1.21\\
                \fiveofive \hspace{0.02cm} & 2019-05-30 & 14 57 15.03 & $-$35 43 50.64 & 18.52 & $1\farcs 0$,$0\farcs 9$,$0\farcs 9$ & 4$\times$/800/750/800 & 0.66/0.66/0.66 &1.5 & 1.22/1.22/1.39&1.18\\           
                \gqc \hspace{0.02cm} & 2019-06-25 & 15 49 13.30 & $-$35 39 11.79 & 18.37 & $1\farcs 0$,$0\farcs 9$,$0\farcs 9$ & 4$\times$/800/750/800 & 1.98/1.965/1.965 &1.5 &  2.6/2.2/2.0 & 1.05\\                 \midrule
        \end{tabular}
        \label{table:observation}
\end{scriptsize}
\end{table*}
\endgroup

\subsection{Data reduction}

Data reduction was carried out using EsoReflex version 2.9.1 \citep[][]{freudling13} and through X-Shooter pipeline version 3.3.5 \citep{modigliani11}. This software has automated the data-reduction process and produces the spectra corresponding to each X-Shooter arm separately using the raw data files. Reducing a science frame requires the following calibration frames: pinhole arc-lamp format-check frames, biases and dark frames for UVB and VIS arms, pinhole continuum-lamp order definition, flat fields, multi-pinhole arc-lamp, and flexure compensation frames. It is noteworthy that flexure compensation frames are optional for more accurate wavelength calibrations. For a detailed description of how each module operates, the reader is referred to the X-Shooter pipeline manual published by ESO.

\subsection{Analysis method}

The spectra extracted from EsoReflex were then analyzed by means of the Image Reduction and Analysis Facility (IRAF)\footnote{IRAF is distributed by the National Optical Astronomy Observatory, which is operated by the Association of the Universities for Research in Astronomy, inc. (AURA) under cooperative agreement with the National
Science Foundation.} and our own software developed in IDL environment\footnote{IDL (Interactive Data Language) is a registered trademark of Harris Corporation.}. Telluric lines were removed using the IRAF telluric package. The flux was corrected for slit losses by dividing the stare mode observations by nodding mode observations. At this stage, the flux was compared with the available photometric data in the literature and good agreement between the two was achieved for each target, within the errors of photometry and flux calibration of the spectra, which is on the order of 15-20\%. The IRAF environment was also used to obtain the physical parameters of each target: SpT, effective temperature ($T_{\rm eff}$), radial velocity (RV), and EWs of spectral lines. We also determined the same physical parameters plus the RV, the projected rotational velocity ($v\sin i$), and the surface gravity ($\log g$) with the version of the code ROTFIT adapted to X-Shooter spectra \citep[][]{frasca17}. We refer the reader to this latter paper for details on this analysis code. In this way, we were able to double-check or revise our final results. In the following paragraph, we give a brief explanation of the general procedure we use for  characterizing the targets. The results from ROTFIT are reported separately in Table \ref{rotfit}.    

To estimate  SpTs for our objects, we first compared the spectrum of each object with a library of stars and brown dwarfs with known parameters observed formerly by X-Shooter \citep[][]{manara13,manara17}. As the spectra are taken with the same instrument and setup, their direct comparison allows us to evaluate SpT and extinction. We then took the average over the spectral indices reported in \citet{riddick2007} to quantitatively determine the  SpTs of our objects. The fitting indices in this paper all belong to the VIS portion of the spectrum, with an error of 0.5 subclasses. Once the SpT was determined, the effective temperature was derived from the calibration of \citet{pm13}\footnote{ \url{http://www.pas.rochester.edu/~emamajek/EEM_dwarf_UBVIJHK_colors_Teff.txt}}. 
We  used the bolometric correction relation proposed by these latter authors to evaluate the luminosity in both \textit{V} and \textit{J} bands and radius of candidates according to their observed parallax and magnitudes. The luminosity of the objects included in Table \ref{stellarprop} was calculated by integrating the flux of X-Shooter spectra and the fitted BT-Settl model, which was double-checked by the luminosities obtained in \textit{V} and \textit{J} bands as explained earlier. Radial velocities (RVs) were determined by cross-correlating portions of the targets' VIS spectra free from emission lines with a template spectrum with the same SpT and low $v\sin i$. All the RVs reported in this paper are extracted from the visual part of the spectrum only, because it has the highest resolution and/or signal-to-noise ratio (S/N) compared to UVB and NIR spectra. The RVs are calculated with the IRAF task {\sc fxcor} and the code ROTFIT, which adopts synthetic BT-Settl spectra \citep[][]{allard12} as templates. The EWs are measured either by fitting a Gaussian to the lines if the line is sharp enough, or direct integration of the pixel values between two marked pixels. A considerable amount of extinction
is not expected for any of our targets, except \gqc. We adopt a zero extinction for cases in which unphysical negative extinctions or extinctions lower than the predicted error (0.5 mag) were achieved\footnote{$A_{v}$ is not determined by ROTFIT, which analyzes the continuum-normalized target spectrum.}. 

Besides these two methods, for a final consistency check, we also fitted the spectrum of each target to the BT-Settl theoretical spectra models \citep[][]{allard12}. These grids of stellar and substellar atmosphere models confirm our conclusions as they estimate the  $T_{\rm eff}$ and $\log g$ of our candidates. To evaluate the age and mass of the targets, we calculated their luminosity and radius first, and then used the available isochrones consistent with these parameters: MESA Isochrones and Stellar Tracks \citep[MIST][]{paxton15,choi16,dotter16} and stellar models generated by \citet{baraffe15}. At this point, we could once again check the  $T_{\rm eff}$ and $\log g$ of our candidates. All the adopted stellar models have solar metallicities.

\section{Results}
\label{results}

In this section, we report the result of our analysis with IRAF and ROTFIT. The kinematic properties of GQ Lupi b are reported in \citet{Neuhauser2008} and \citet{schwarz16}. It is noteworthy that \threetn \hspace{0.02cm} is a possible member of the $\beta$-Pictoris MG, while the other four candidates are located in various clouds of the Scorpius-Centaurus association (hereafter Sco-Cen). Sco-Cen is the closest OB association to the Sun, consisting of three main clouds and some areas with a diffuse kinematical population \citep[][and references therein]{damiani19}. The central stars of these candidates  are acknowledged as members of the association according to \citet{rizzuto11}. For all the candidates, we also checked the astrometric excess noise and blue-red (\textit{BP-RP}) excess factors reported by \textit{Gaia} DR2 to check for signs of binarity. None of the targets show a significant \textit{BP-RP} excess factor ($\sim$ 2-3 addressing a source  heavily affected by noise or a highly populated region), while\hspace{0.02cm} a significant astrometric excess noise ($>$ 1) was reported for \fiveothree \hspace{0.02cm} and
\fiveofive. This latter has been confirmed by a considerable value of astrometric-excess-noise-sigma ($>$ 2) reported for the two objects. The spectra acquired for these two candidates are rather noisy compared to the other three and were more challenging to analyze. 

The excess noise takes the modeling errors into account \citep{lindergen12} and a value close to zero indicates an astrometrically well-behaved single source. However, not all sources behave exactly as the best-fitting standard astrometric model. For example, mid-late M-type members of the Upper Scorpius sub-association have high astrometric excess noise because they are intrinsically faint and the five-parameter fit is therefore imperfect. We therefore investigated the members of Upper Centaurus-Lupus (UCL) and Lower Centaurus-Crux (LCC) studied by \citet{damiani19} to understand whether or not the members of these two sub-associations, to which three of our candidates belong,  also show significant astrometric excess noise. Our studies indicate that among the upper main sequence (UMS) and PMS members of LCC, collectively, 75 out of 90 members have a significant astrometric excess noise, and this number is 597 out of 697 members for UCL. We also calculated the renormalised unit weight error (RUWE) associated to each source which is more reliable as an indicator of the goodness-of-fit statistic than astrometric excess noise \citep{lindergen18}.                

For two of our targets, namely \threetn \hspace{0.02cm} and \fiveofive \hspace{0.02cm}, we could not measure the $EW_{\ion{Li}{i}}$ because of non-detection of the lithium line, and therefore we only report an upper limit. For this purpose, a three-sigma upper limit on the flux of the lithium line is calculated based on the spectra of the  objects \citep{cayrel1988}:

\begin{equation}
dEW = 3 \times 1.06 \sqrt{(FWHM) \hspace{0.02cm} dx} \hspace{0.02cm}/\hspace{0.02cm} (S/N) ,   
\end{equation}

\noindent in which FWHM is the full width at half maximum, S/N is the signal-to-noise ratio, and the bin size (dx) is 0.2 \AA \hspace{0.02cm} for the VIS arm. In Fig. \ref{ew-lith}, we indicate the upper limits of $EW_{\ion{Li}{i}}$ for the aforementioned candidates as downward arrows.

\subsection{\threetn}  

\begin{figure}[ht!]
    \subfloat[]{
        \begin{minipage}[b]{.5\textwidth}
        \centering
        \includegraphics[width=9cm]{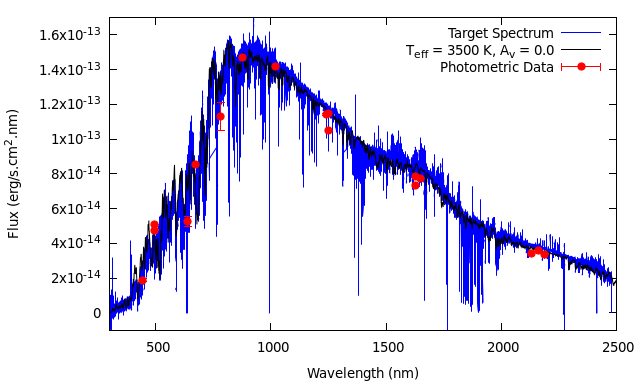}    
\label{329-a}        
        \end{minipage}
        }\quad  
  \subfloat[]{
        \begin{minipage}[b]{
           0.5\textwidth}
        \centering
        \includegraphics[width=9cm]{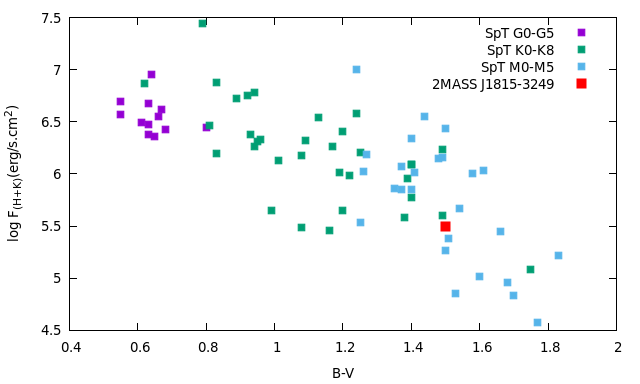}
\label{329-b}    
    \end{minipage}}
    \caption{\protect\subref{329-a} Spectrum of \threetn \hspace{0.02cm}   (blue)  together with a BT-Settl model (black) fitted to the spectrum assuming zero extinction. \protect\subref{329-b} Total excess emission in \ion{Ca}{ii} H and K vs. color index B-V of objects studied by \citet{lopez10} and \citet{martinez-ar2011} including \threetn \hspace{0.02cm}, shown as a red square.}
\end{figure}

The spectrum of \threetn \hspace{0.02cm} (Fig. \ref{329-a}) displays clear features of an M2--3 type star, which is reported in Table \ref{stellarprop} in addition to its other stellar parameters. For this star, the measured $v\sin i$ with ROTFIT is lower than 8 km/s. As the minimum detectable $v\sin i$ at this resolution is 8 km/s \citep[][]{frasca17}, we report only an upper limit of 8 km/s. The \ion{Li}{i} line was not detected in the spectrum. However, this is not conclusive for determining the age of the object as no lithium is expected for a M2--3 star even at the young age of $\beta$ Pictoris MG (Fig. \ref{ew-lith}). Regarding the chromospheric activity, the $H_{\alpha}$ line was found to be in absorption and moderate \ion{Ca}{ii}-IRT activity was detected. Regarding the physical association with the central star, V4046 Sgr was measured to have a RV = $-$6.94$\pm$0.16; its center of mass RV was obtained through orbital solution by \citet{quast2000}; and its low-mass companion, GSC 7396-00759, has RV = $-$6.10$\pm$0.5 \citep[][]{sissa18}. The values of RV derived by us for 2MASS J1815-3249 with the two adopted procedures (Tables 3 and 4) agree with each other within the errors, but are significantly different from the barycentric velocity of V4046 Sgr and the RV of GSC 7396-00759. This argues against a physical association of this star  with the stellar system; for a summary of their collective kinematic properties see Table \ref{table:kinematic}. The chromospheric activity is evident as a small filling in the cores of \ion{Ca}{ii}-IRT lines and emission above the continuum level in the cores of the \ion{Ca}{ii} H and K lines. We measured the flux in the H and K lines, integrating the emission features arising from the subtraction of the inactive template produced by the BT-Settl spectrum with $T_{\rm eff}=3500$\,K.
We compared the H and K flux of \threetn \hspace{0.02cm} with that of late-type active stars investigated by \citet{lopez10} and \citet{martinez-ar2011}. As can be seen in Fig. \ref{329-b}, the position of our target is well below the upper branch of the very active, young M-type stars and lies in the region occupied by older objects ($Age \geq 150$\,Myr). An age older than that of the $\beta$-Pictoris MG is also suggested by the CMD presented in Fig. \ref{cmd}, which exhibits \threetn \hspace{0.02cm} on the MS branch, unlike the other young candidates discussed in this work, which lie significantly higher.

The X-ray luminosity as derived from the observed X-ray flux \citep[][]{rosen16} coupled with our adopted stellar parameters yields $\log (L_X / L_{bol})=-4.14$. This value is consistent with those of Hyades stars of similar colors, and well below those of members of younger clusters or groups such as the Pleiades and the
$\beta$-Pictoris MG \citep[][]{song2008}, indicating a most probable age of several hundred million years.

The age inconsistency with the $\beta$-Pictoris MG, which is supported by the position of \threetn\  on the CMD, a low level of magnetic activity, and the absence of lithium absorption, together with the discrepant RV with respect to that of the central star imply that this target is not a legitimate member of the group and is significantly older than expected.

\subsection{\foreifor}

\begin{figure}[ht!]
        \includegraphics[width=\columnwidth]{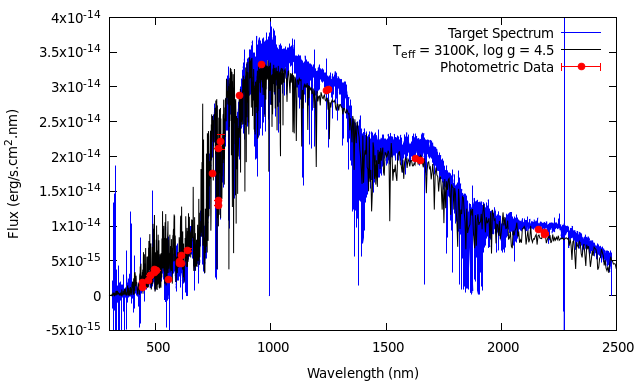}
    \caption{Spectrum of \foreifor \hspace{0.02cm}  (blue) together with the available photometric data (red dots) in the literature and the fitted BT-Settl model (black).}
    \label{fig:484}
\end{figure}

\foreifor \hspace{0.02cm} has already been identified as a PMS member of UCL \citep{damiani19}. The assessed age for the UCL sub-association was reported to be 15$\pm$3 Myr by \citet{pm16}, and therefore we adopt the same value as a first guess for our target. We also assume the same age for the central star which is an F4V star; for a detailed study, the reader is referred to \citet{pm12}. The physical stellar parameters and ROTFIT results derived from the X-Shooter spectrum are  in good agreement with each other, and are displayed in Tables \ref{stellarprop} and \ref{rotfit}, respectively. The BT-Settl synthetic template, which is overplotted on the X-Shooter target spectrum in Fig. \ref{fig:484}, confirms the values of $T_{\rm eff}$ and $\log g$ derived with ROTFIT. The fluxes derived from literature photometry, which are overplotted as red dots in Fig. \ref{fig:484}, support the flux calibration we performed. The EW of the lines relevant for assessing chromospheric activity are incorporated in Table \ref{EW}. The EW of $H_{\alpha}$ and other hydrogen lines in emission plus $EW_{Li}$ indicate a young and active star, meeting our preliminary expectations. These values also demonstrate that the object is not actively accreting matter from a circumstellar disk, as none of the measured hydrogen lines are sufficiently intense or wide. The central star, HIP 74865, which we initially surmised to be physically bound with \foreifor \hspace{0.02cm}, has an RV = 2$\pm$0.3 \citep[][]{chen11} and for the companion we have RV = 1.4$\pm$2.4. This confirms the two stars share a significant similarity in kinematic properties; see Table \ref{table:kinematic} for a summary of their kinematic properties. Based on the stellar parameters of \foreifor, \citet{baraffe15} models suggest an age of 7.4$\pm$0.5 Myr for this object which is younger than the age of the UCL sub-association by a factor of two. In a recent paper by \citet{asensio19}, who investigated an Upper Scorpius-Centaurus member and its companions, namely HIP 79124ABC, the same problem is discussed. Although the components of  this stellar system are expected to be coeval, both of the low-mass companions were estimated to have half the age of  the central star. \citet{feiden16} also pointed out that cooler K and M stars located in young stellar associations are reported to be younger than hotter A, F, or G stars belonging to the same association by a factor of two. This problem is probably due to neglect of the magnetic fields in stellar evolution isochrones which in turn fail to reproduce the age of low-mass PMS stars accurately \citep{asensio19,feiden16}. Therefore, by adopting the most appropriate stellar evolution models tailored for ages beyond 10Myr, we expect \foreifor \hspace{0.02cm} to have an age consistent with UCL. All aspects considered, we conclude there is a high chance that this candidate is physically associated with HIP 74865, offering evidence of a new triple system. The projected separation
of about 11100 au also supports the idea of this configuration, indicating a physically bound system.
  
\subsection{\fiveothree}

\begin{figure}[ht!]
        \includegraphics[width=\columnwidth]{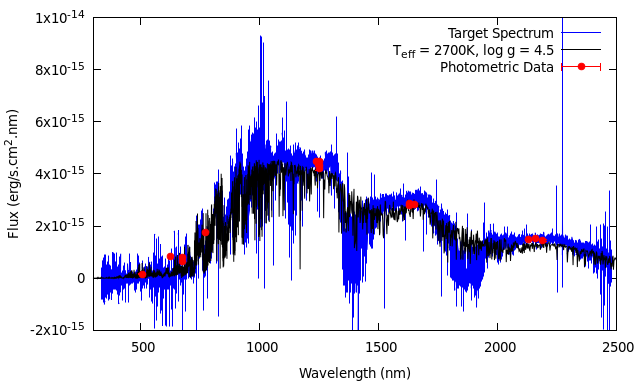}
    \caption{Spectrum of \fiveothree \hspace{0.02cm} (blue) with the photometric data (red dots) and the BT-Settl model (black) that confirms the estimated $T_{\rm eff}$ and $\log g$.}
    \label{fig:503}
\end{figure}

\begin{figure}[ht!]
    \subfloat[]{
        \begin{minipage}[b]{.5\textwidth}
        \centering
        \includegraphics[width=9cm]{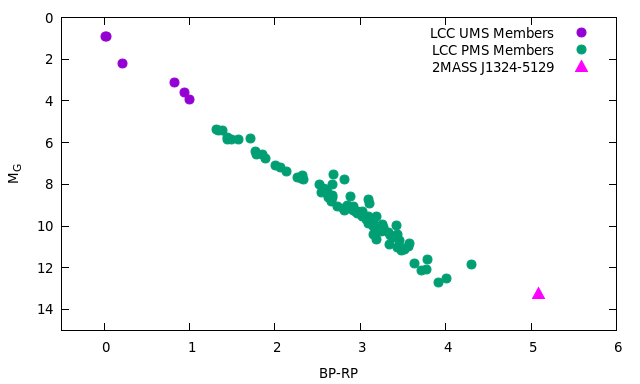}    
\label{lcc-cmd}        
        \end{minipage}
        }\quad  
  \subfloat[]{
        \begin{minipage}[b]{
           0.5\textwidth}
        \centering
        \includegraphics[width=9cm]{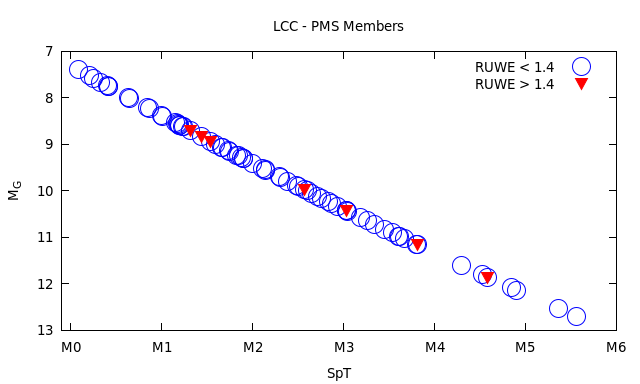}
\label{lcc-ruwe}    
    \end{minipage}}
    \caption{\protect\subref{lcc-cmd} UMS and PMS members of LCC \citep{damiani19} with our target in the same sub-association, \fiveothree \hspace{0.02cm}.  \protect\subref{lcc-ruwe} RUWE of M-type members of LCC. The blue circles display members with RUWE indices below the threshold of RUWE $\sim$ 1.4, while the red triangles represent members with RUWE $>$ 1.4 which refer to nonsingle or problematic objects for the astrometric solution.}
\end{figure}

This target is situated in the LCC sub-association of Sco-Cen, with a projected separation of 15500 au from the central star, and has not  yet been listed as a member of Sco-Cen in the literature. The \ion{Li}{i} line was found to be in absorption and various hydrogen lines are in emission, as indicated in Table \ref{EW}. The $H_\alpha$ line is wider than expected as it is apparently contaminated by cosmic ray hits on the detector. The  spectrum of this target  shows features of an object with SpT of M6.5$\pm$0.5 subclasses (Fig. \ref{fig:503}). \citet{reyle18} evaluated the SpT of  this object as M7 which is in good agreement with our estimation. We placed this object along with UMS and PMS members of LCC on a CMD (Fig. \ref{lcc-cmd}) that shows \fiveothree \hspace{0.02cm} is fainter than the studied sample by \citet{damiani19}. As we pointed out at the beginning of this section, most of the objects in this sample have a significant astrometric excess noise. We therefore calculated only the RUWE of the LCC M-type members to make a comparison between the fainter objects of the sample and our target. As displayed in Fig. \ref{lcc-ruwe}, a few of these LCC members have a significant RUWE, above the threshold of 1.4, implying a nonsingle object or nonconsistent astrometric solution with the observations. The SpT of the members is estimated based on their $M_{G}$ \citep[][]{kiman19}. The RUWE of our object is 1.104, which is not significantly higher than 1 and confirms the source is well-behaved.

Very recently, the RV of HIP 65426 and its companion HIP 65426 b were revisited (Petrus et al. 2020, submitted). According to these new evaluations, RV=12.2$\pm$0.3\,km/s for HIP 65426 and RV=26$\pm$15\,km/s for HIP 65426 b. These values are well within the RV range we report for \fiveothree \hspace{0.02cm} which is 15.29$\pm$5.75\,km/s (although with large error bars). While displaying the typical features of an active young stellar object (YSO) and consistent RV with the central star, the slightly discrepant kinematic properties of \fiveothree \hspace{0.02cm} and HIP 65426 argue against their physical association (see Table \ref{table:kinematic}). As in \foreifor, \fiveothree \hspace{0.02cm} is also not accreting matter from the central star according to Fig. \ref{fig:manara}. The position of \fiveothree \hspace{0.02cm} on the CMD in Fig. \ref{cmd} confirms its status as a PMS star. We checked the membership of this target through BANYAN $\Sigma$ \citep[][]{gagne18} and the probability of \fiveothree \hspace{0.02cm} being a member of LCC is above 94\%. We therefore conclude that \fiveothree \hspace{0.02cm} is a new PMS member of LCC. The age of 16.0$\pm$2.2\,Myr that we derive from the HR diagram and the \citet{baraffe15} evolutionary tracks is in perfect agreement with the age reported by \citet{pm16} for the LCC sub-association (16.0$\pm$2.2\,Myr). The mass that we derive with the same models, $M_{\star}=0.04\,M_{\sun}$, confirms 2MASS J1324-5129 as a substellar object.

\subsection{\fiveofive}  
\label{fiveofive}

\fiveofive \hspace{0.02cm} is situated in UCL sub-association in Sco-Cen, the same as \foreifor \hspace{0.02cm}, and has not been confirmed as a member of the association in the literature. As discussed above, a physical association with HIP 73145 at 280\arcsec \hspace{0.02cm} is unlikely, because of the slightly discrepant kinematic parameters. In order to have a full 
comparison of the kinematics, we derived the RV of the central star\footnote{The RV quoted in SIMBAD for HIP 73145 at the time of writing is not a true measurement, but rather the kinematic RV expected for a member of the association as derived in \citet{madsen02}.} using
the reduced FEROS and HARPS spectra of HIP 73145 available on ESO archive\footnote{ \url{archive.eso.org}  }.

\begin{figure}[ht!]
        \includegraphics[width=\columnwidth]{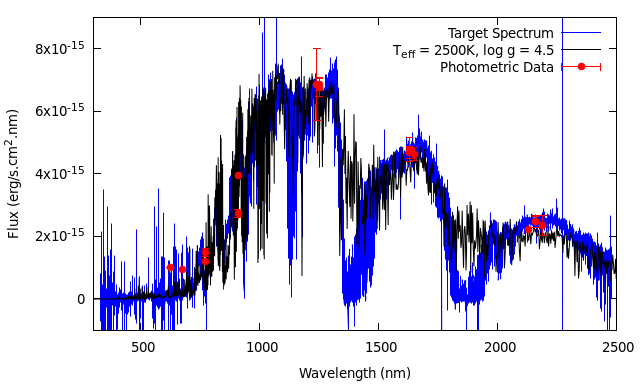}
    \caption{Spectrum of \fiveofive \hspace{0.02cm} (blue) with the available photometric data (red dots) from the literature for checking the flux calibration, and the BT-Settl model (black) for double-checking the estimated $T_{\rm eff}$ and $\log g$.}
    \label{fig:505}
\end{figure}

According to \citet{riddick2007} SpT indices, \fiveofive \hspace{0.02cm} is an M8 ($\pm$0.5 subclasses) dwarf. Clear signs of activity, such as Balmer emission lines, are present in its spectrum. The kinematic properties of the target are compatible with Sco-Cen membership (with its UCL membership probability higher than 97\% according to BANYAN). However, there is one crucial element missing in the spectrum of this   target
considering the general expectations from a Sco-Cen object of this spectral type: \ion{Li}{i}. This nondetection could be due to the fast rotational velocity $v\sin i$ = 62.0$\pm$10\,km/s of this  target or the low S/N ($\sim$ 5) in that region. In this scenario, which is the most likely explanation, \fiveofive \hspace{0.02cm} is a member of UCL with an estimated age of 17 Myr according to \citet{baraffe15} models. We estimate an upper limit of $dEW_{\ion{Li}{i}} = 390 \hspace{0.1cm} m\AA$ for this target (indicated as a downward arrow in Fig. \ref{ew-lith}). Its estimated $T_{\rm eff}$ and $\log g$ are confirmed by fitting a BT-Settl model with the same qualities to its spectrum (Fig. \ref{fig:505}).

Another possibility is that the \ion{Li}{i} line is absent in the spectrum of this  target. In this case, it is possible that the object is older than the UCL sub-association. By comparing the models of \citet{baraffe15} and empirical results of Li distribution
in the $\sim$ 50 Myr-old IC 2391 open cluster \citep[][]{barrado}, we reach the conclusion that Li nondetection implies \fiveofive \hspace{0.02cm} is older than 80 Myr. 


An age of 80-150 Myr (e.g., similar to the Pleiades) can still be compatible with a high activity level and fast rotation.
Alternatively, activity and rotation might be induced by a close companion through tidal locking.
Some indications of an unresolved binary are suggested by the astrometric excess noise reported for this object, which is equal to 0.99 mas with a significant astrometric-excess-noise sigma of 3.29.
However, a very close, tidally locked binary is not expected to cause significant astrometric effects.
In any case, while the spectral lines are very broad, there are no indications of blending of additional components. Therefore, the possibility of a binary object remains speculative at this stage.

\begin{figure}[ht!]
    \subfloat[]{
        \begin{minipage}[b]{.5\textwidth}
        \centering
        \includegraphics[width=9cm]{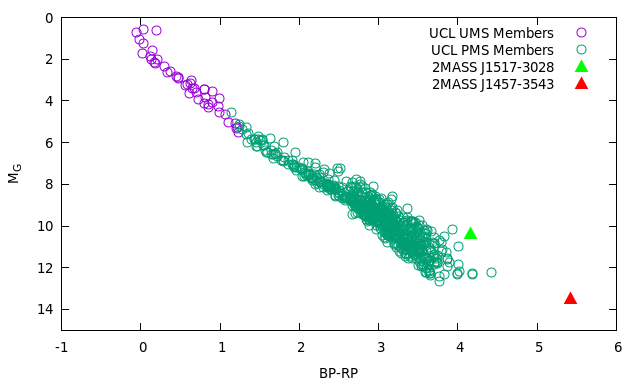}    
\label{ucl-cmd}        
        \end{minipage}
        }\quad  
  \subfloat[]{
        \begin{minipage}[b]{
           0.5\textwidth}
        \centering
        \includegraphics[width=9cm]{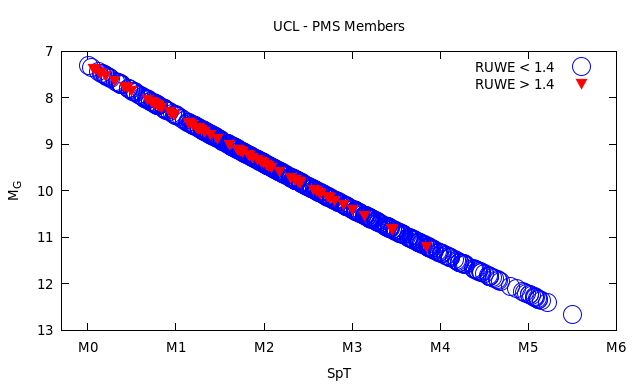}
\label{ucl-ruwe}    
    \end{minipage}}
    \caption{\protect\subref{ucl-cmd} UMS and PMS members of UCL \citep{damiani19} with our targets located in the same sub-association, namely, \fiveofive \hspace{0.02cm} and \foreifor. \protect\subref{ucl-ruwe} RUWE of M-type members of UCL. The blue circles indicate the members with RUWE indices below the threshold of 1.4, while the red triangles represent members with RUWE $>$ 1.4. Only 68 of 588 objects have a significant RUWE.}
\end{figure}

We also put \fiveofive \hspace{0.02cm} on a CMD with other UCL members \citep{damiani19}, plus our other target \foreifor \hspace{0.02cm} located in the same sub-association (Fig. \ref{ucl-cmd}). This CMD re-confirms \foreifor \hspace{0.02cm} as a PMS member of UCL, and also shows that \fiveofive \hspace{0.02cm} is fainter than the UCL members studied by \citet{damiani19}. We calculated RUWE of the M-type objects in this sample and found that 68 out of 588 of them have RUWE $>$ 1.4 (Fig. \ref{ucl-ruwe}). The  SpTs of the members are calculated according to their $M_{G}$ \citep[][]{kiman19}. The RUWE of \fiveofive \hspace{0.02cm}, although intrinsically fainter than the members, is equal to 1.174 that is below RUWE $\sim$ 1.4, further  decreasing the chance of our target being a nonsingle object. 


We also checked whether the large $v\sin i$ of \fiveofive \ is compatible with its SpT. According to \citet{jeffers18}, there is a correlation between the $H_{\alpha}$ activity and $v\sin i$ of M-type dwarfs: the trend shows earlier SpTs have lower rotation and activity levels, while later SpTs are mostly fast rotating and chromospherically active stars. The studied sample in \citet{jeffers18} is not sufficiently large to verify such a correlation, but it seems \fiveofive \hspace{0.02cm} falls in the latter category of late M-type dwarfs which are fast rotating and active. However, the reported $v\sin i$ = 62.0\,km/s is higher than the values found in the Carmencita catalog for the same SpT in the same paper. We note that higher values were obtained for earlier and later M-type dwarfs in the same catalog. A larger sample was studied by \citet{konopacky12}, who focused on determining the rotational velocities of very late-type, low-mass individual binary components. According to this latter work, rotational velocities as high as the value found for \fiveofive \hspace{0.02cm} (of M8 SpT or later) had been identified before in the literature. More importantly, \citet{dahm12} studied 94 members of Upper Scorpius (USC) two of which have an M8 SpT, with measured projected rotational velocities of 22.70$\pm$4.47 and 60.94$\pm$3.03\,km/s. The latter, which has a  similar rotational velocity to our target, is identified as a fast rotator. In total, 4 out of 38 objects with SpTs between M4 and M8 possess rotational velocities higher than 30\,km/s, which confirms that the  rotational velocity of our target is not uncommon among very young objects.

From $v\sin i$ and stellar radius, we estimated a rotation period of 0.2 days for \fiveofive. Recent studies on M dwarfs belonging to Pleiades and Praesepe clusters \citep[][]{somers17} and low-mass stars residing in USC and $\rho$ Ophiuchus \citep[][]{rebull18} suggest that rotation periods much shorter than 1 day are frequent for the
lowest mass objects ($\leq$ 0.1 \Msun).
Finally, the position of \fiveofive \hspace{0.02cm} on a CMD is well above the main sequence (Fig. \ref{cmd}) and along the sequence of Sco-Cen members. However, it is also possible that the overluminosity is due to binarity.

We conclude that the most likely scenario explaining the spectrum of \fiveofive \hspace{0.02cm} is that the target is a young (17.75$\pm$4.15 Myr) and active late-type M dwarf and a new member of  the UCL. Nevertheless, both of the other alternatives for the target, namely (i) a highly active, fast-rotating young object with an approximate age of 100 Myr, or (ii)  a close binary, remain viable. As we cannot exclude that the observed spectrum is that of a spectroscopic binary, the stellar parameters we derive from it must be taken with caution, as they are only reliable if the object is a single star. However, it is noteworthy that if our evaluations  of \fiveofive \hspace{0.02cm} as a single object are accurate, this target might be a very wide companion of HIP 73145. The RV=4.01$\pm$8.87\,km/s and age (17.75$\pm$4.15 Myr) of \fiveofive\  are consistent with those of HIP 73145 (RV=3.8$\pm$1.6\,km/s, age$\sim$15 Myr), although their inconsistent parallax and the wide projected separation argues against their physical association.

Finally, as the age of this object is highly uncertain, its mass cannot be unambiguously estimated.
We point out that \fiveofive \hspace{0.02cm} should be substellar ($<$ 0.069 \Msun) if the object is younger than 144.5 Myr. Masses of 0.03 and 0.06 \Msun are derived for 15.5 Myr (Sco-Cen age) and 120 Myr (Pleiades age), respectively.

\begin{table*}
        \centering
        \caption{Physical stellar parameters of candidates. Except for \threetn \hspace{0.02cm}, mass, age, and $\log g$ are selected from the closest matches offered by \citet{baraffe15} models. The same quantities are extracted from MIST isochrones for \threetn \hspace{0.02cm}.}
        \begin{minipage}{\textwidth}
        \centering
        \begin{tabular}{lccccccccr} 
                \midrule
                Name & SpT & $T_{\rm eff}$ & $A_v$ & RV & $L_{\star}$ & $R_{\star}$ & $M_{\star}$  & Age & $\log g$ \\
                 & & (K) & (mag) & (km/s) & (\Lsun) & (\Rsun) & (\Msun) & (Myr) & \\
                \midrule
                \threetn \hspace{0.02cm} & M3$\pm$0.5 & 3410 & 0 & $-$17.7$\pm$2.4 & 0.03 & 0.5 &  0.3 & $>$ 150 & 4.95 \\
                \foreifor \hspace{0.02cm} & M4.5$\pm$0.5 & 3100 & 0 & 1.2$\pm$6.9 & 0.018 & 0.47 & 0.11 & 7.9 & 4.13 \\ 
                \fiveothree \hspace{0.02cm} & M6.5$\pm$0.5 & 2710 & 0 & 15.3$\pm$5.8 & 0.0024 & 0.22 & 0.04 & 18.2 & 4.35 \\
                \fiveofive \hspace{0.02cm} & M8$\pm$0.5 & 2500 & 0 & 4.0$\pm$8.9 & 0.002 & 0.24 & 0.02 & 14.09 & 4.00 \\         
                \gqc \hspace{0.02cm} & M4$\pm$0.5 & 3190 & 1.0$\pm$0.5 & $-$1.3$\pm$4.0 & 0.07 \footnote{This object is sub-luminous. The value reported here is the object's corrected luminosity obtained by \citet{alcala20}.} & 0.21 &  0.15 & 2 & 3.78 \\
                \hline
        \end{tabular}
        \end{minipage}
        \label{stellarprop}
\end{table*}

\begin{table*}
        \centering
        \caption{ROTFIT results acquired for the targets. The estimated ages are concluded from \citet{baraffe15} models.}
        \begin{tabular}{lccccc} 
                \midrule
                Name & $T_{\rm eff}$ & $\log g$ & RV & $v\sin i$ & Age \\
                & (K) &  & (km/s) & (km/s) & (Myr)  \\
                \midrule
                \threetn \hspace{0.02cm} & 3562$\pm$30 & 4.68$\pm$0.14 & $-$20.1$\pm$2.0 & $<$ 8.0 & $>$ 150 \\
                \foreifor \hspace{0.02cm} & 3077$\pm$22 & 4.49$\pm$0.21 & 1.4$\pm$2.4 & 26$\pm$6 & 7.4$\pm$0.5 \\         
                \fiveothree \hspace{0.02cm} & 2646$\pm$50 & 4.0$\pm$0.1 & 17.9$\pm$3.0 & $<$ 8.0 & 16$\pm$2.2 \\
                \fiveofive \hspace{0.02cm} & 2635$\pm$80 & 4.30$\pm$0.17 & 8.0$\pm$7.8 & 62.0$\pm$10.0 & 17.75$\pm$4.15 \\               
                \gqc \hspace{0.02cm} & 3230$\pm$101 & 3.74$\pm$0.23 & $-$2.0$\pm$2.8 & 13.0$\pm$6.0 & 2.75$\pm$0.75  \\                      \midrule
        \end{tabular}
        \label{rotfit}
\end{table*}

\begin{table*}
        \centering
        \caption{Kinematic properties of our targets and their associated stellar systems. For a discussion on the adopted kinematic properties of \gqc, \hspace{0.02cm} the reader is referred to \citet{alcala20}; we have adopted the parallax of the central star for \gqc \hspace{0.02cm} because of its highly uncertain measured kinematic features as the only accreting object among the targets.}
        \begin{minipage}{\textwidth}
        \centering
        \begin{tabular}{lccccccr} 
                \midrule
                Name & parallax & $\mu_\alpha$  & $\mu_\delta$ & RV \\
                 &  (mas) & (mas/yr) & (mas/yr) & (km/s) \\
                \midrule
                \threetn \hspace{0.02cm} & 13.12$\pm$0.054 & 1.07$\pm$0.095 & $-$52.74$\pm$0.078 & $-$20.1$\pm$2.0 \\
                V4046Sgr & 13.81$\pm$0.064 & 3.49$\pm$0.11 & $-$52.75$\pm$0.087 & $-$6.94$\pm$0.16 \\           
                GSC 7396-00759 & 13.99$\pm$0.052 & 3.08$\pm$0.10 & $-$52.64$\pm$0.08 & $-$6.10$\pm$0.5 \\    
                \midrule
        
                \foreifor \hspace{0.02cm} & 8.16$\pm$0.11 & $-$21.67$\pm$0.21 &       $-$28.31$\pm$0.18 & 1.4$\pm$2.4 \\
                HIP 74865 & 8.09$\pm$0.061 & $-$21.07$\pm$0.11 & $-$28.42$\pm$0.10 & 2.0$\pm$0.3 \\                
                                \midrule
                
                \fiveothree \hspace{0.02cm} & 8.01$\pm$0.35 & $-$31.85$\pm$0.53 & $-$17.07$\pm$0.44 & 15.29$\pm$5.75 \\
                HIP 65426 & 9.16$\pm$0.062 & $-$34.25$\pm$0.10 & $-$18.81$\pm$0.093 & 12.2$\pm$0.3 \\               
                \midrule
                \fiveofive \hspace{0.02cm} & 9.86$\pm$0.42 &    $-$28.68$\pm$0.68 &       $-$27.3$\pm$    0.65    & 8.0$\pm$7.8\\
                HIP 73145 & 7.48$\pm$0.20 &     $-$23.35$\pm$0.26 & $-$24.94$\pm$0.30 & 3.8$\pm$1.6\\
                \midrule
                \gqc \hspace{0.02cm} & 6.59$\pm$0.05 \footnote{Considering that the astrometric parameters of GQ Lup C are somewhat biased by illumination effects (due to the presence of a disk and likely outflows), the parallx of GQ Lup has been adopted for this target; see \citet{alcala20} for more information.} & $-$14.81$\pm$0.97 & $-$21.95$\pm$0.65 & $-$2.0$\pm$2.8 \\
                 GQ Lup & 6.59$\pm$0.05 & $-$14.26$\pm$0.01 & $-$23.6$\pm$0.07 & $-$3.6$\pm$1.3 \\
                 GQ Lup B & 7.2$\pm$2.1 & - & - & 2.0$\pm$0.4 \\
                 \midrule
        \end{tabular}
\label{table:kinematic}
\end{minipage}
\end{table*}

\begin{table*}
        \centering
        \caption{Photometry of our targets.}
        \begin{tabular}{lccccccr} 
                \midrule
                Name & \textit{BP-RP} & \textit{J} & \textit{H} & \textit{Ks} & \textit{G}  \\
                & (mag) & (mag) & (mag) & (mag) & (mag) \\
                \midrule
                
                \threetn \hspace{0.02cm} & 2.25 & 11.07 & 10.44 & 10.2 & 13.6  \\
                \foreifor \hspace{0.02cm} & 3.18 & 12.54 & 11.93 & 11.64 & 15.8 \\               
                \fiveothree \hspace{0.02cm} & 3.87 & 14.59 & 14.04 & 13.63 & 18.71 \\
                \fiveofive \hspace{0.02cm} & - & 14.13 & 13.5 & 13.1 & 18.52 \\              
                \gqc \hspace{0.02cm} & 2.75 & 14.85 & 14.08 & 13.82 & 18.37  \\                     \midrule
        \end{tabular}
\label{table:colors}
\end{table*}

\begin{table*}
        \centering
        \caption{Equivalent width of the relevant lines indicating chromospheric and accretion tracers for our candidates.}
\begin{minipage}{\textwidth}
\begin{tiny}
\centering
        \begin{tabular}{lcccccccr} 
                \midrule
                Name & $EW_{\ion{Li}{i}}$  & $EW_{H_{\alpha}}$ & $EW_{H_{\beta}}$ & $EW_{H_{\gamma}}$   \\
                & (nm) & (nm) & (nm) & (nm)  \\
                
                \midrule
                
                \threetn \hspace{0.02cm} & $<$ 0.0036 \footnote{\label{star}Three-sigma upper limits (see the text for further explanation). The $EW_{\ion{Li}{i}}$ of 2MASS J1549-3539 may be affected by veiling \citep{alcala20}.} & 0.012$\pm$0.001 & $-$0.034$\pm$0.003 & $-$0.013$\pm$0.005  \\
                \foreifor \hspace{0.02cm} & 0.038$\pm$0.008 & $-$1.083$\pm$0.055 & $-$1.019$\pm$0.067 & $-$1.126$\pm$0.201  \\           
                \fiveothree \hspace{0.02cm} & 0.070$\pm$0.020 & $-$1.068$\pm$0.142 & $-$0.659$\pm$0.149 & $-$0.723$\pm$0.129   \\
                \fiveofive \hspace{0.02cm} & $<$ 0.039 \footref{star} & $-$0.983$\pm$0.090 & $-$0.789$\pm$0.166 & $-$0.886$\pm$0.135 \\            
                \gqc \hspace{0.02cm} &  0.035$\pm$0.010 & $-$9.943$\pm$0.252 & $-$4.085$\pm$0.701 & $-$2.722$\pm$0.217   \\                  \midrule
        \end{tabular}
        \label{EW}
        
\end{tiny}
\end{minipage}
\end{table*}

\section{Discussion and Conclusions}
\label{discussion}

\begin{figure}[ht!]
    \subfloat[]{
        \begin{minipage}[b]{.5\textwidth}
        \centering
        \includegraphics[width=9cm]{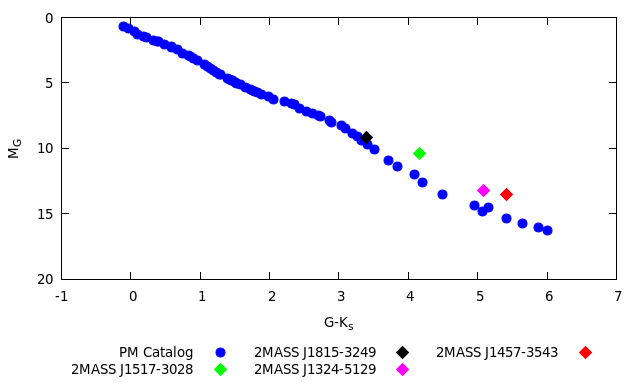}
        \label{cmd}
        \end{minipage}}\quad    
  \subfloat[]{
        \begin{minipage}[b]{
           0.5\textwidth}
        \centering
        \includegraphics[width=9.2cm]{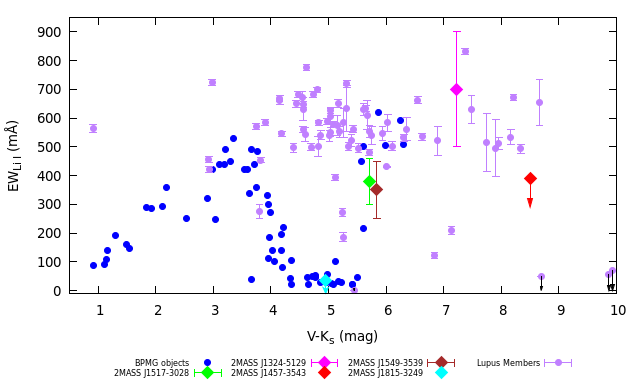}
        \label{ew-lith}
    \end{minipage}}
    \caption{\protect\subref{cmd} CMD of the wide companion candidates
in this work (diamond) together with the objects in \citet[][blue dots]{pm13}. \protect\subref{ew-lith} $EW_{\ion{Li}{i}}$ of our targets (diamonds) plotted with Lupus members \citep{biazzo17} and 66 members of the $\beta$ Pictoris association \citep{messina16}.}
\end{figure}

We analyzed the X-Shooter spectra of probable wide companions of four different stellar systems to determine their physical parameters. \threetn \hspace{0.02cm} was found to be a field object unrelated to the $\beta$-Pictoris MG,  with an estimated age of $>$ 150 Myr, in spite of its parallax and proper motions, which are similar to the bona-fide member V4046 Sgr. The results we obtained for the other three targets located in Sco-Cen are summarized below. All the four wide companion candidates analyzed in this paper are placed on a CMD together with the objects studied in \citet{pm16} (featured as blue dots in Fig. \ref{cmd}). This CMD confirms \fiveothree \hspace{0.02cm} and \foreifor \hspace{0.02cm} as PMS members of Sco-Cen sub-associations. For \fiveofive , considering all the aspects addressed in the previous section, we find that the most likely scenario indicated by the target spectrum is that of a single object and a PMS (evident from Fig. \ref{cmd}) member of UCL. Following this scenario, \fiveofive \hspace{0.02cm} could also be an ultra-wide companion to HIP 73145 considering their consistent ages and RVs, although their discrepant parallax reported by \textit{Gaia DR2} argues against this possibility.

\begin{figure}[!ht]
\centering
\includegraphics[width=9cm]{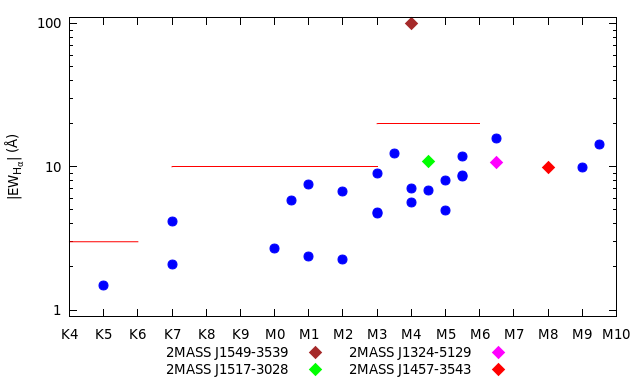}
\caption{$EW_{H_{\alpha}}$ vs. SpT of our targets with $H_{\alpha}$ emission (diamonds) along with the objects studied by \citet[][blue dots]{manara13}. The red horizontal lines represent the thresholds between nonaccreting and accreting objects for different SpT ranges according to \citet{white2003}.}
\label{fig:manara}
\end{figure}

The criteria used for distinguishing between accreting and nonaccreting objects are those proposed by \citet{white2003}, which are based on the EW of H$_{\alpha}$ as a function of SpT and the full width of the line profile at 10\% of the peak. The plot of $EW_{H_{\alpha}}$ against SpT (Fig. \ref{fig:manara}) for our candidates and the T Tauri stars
studied by \citet{manara13} suggests that the only accreting object in our sample is 2MASS J1549-3539, which was fully investigated by \citet{alcala20}. A checklist of the properties of our  targets relevant for their wide-companion membership status is presented in Table \ref{status}.

$EW_{\ion{Li}{i}}$ of $\beta$-Pictoris MG and Lupus members measured by \citet{messina16} and \citet{biazzo17}, respectively, are plotted along with the $EW_{\ion{Li}{i}}$ of our targets; see Fig. \ref{ew-lith} \footnote{The V band magnitude for our program targets was derived from the \citet{pm13} tables (the updated table can be found on-line at \url{http://www.pas.rochester.edu/~emamajek/EEM_dwarf_UBVIJHK_colors_Teff.txt}}. Our only target located in $\beta$-Pictoris MG, namely \threetn, \hspace{0.02cm} which is determined as a field object, is depleted of lithium as expected, taking into account its age (older than 150 Myr). Members of $\beta$-Pictoris MG depleted of lithium pinpoint the age of the association, as lithium is burned rapidly in low-mass stars due to PMS contraction increasing stellar core temperature to as high as $\sim 3 \times 10^6$ K. Subsequently, depending on the mass of the stars, at a certain age, convective mixing depletes the stars of lithium entirely. The lithium gap existing among $\beta$-Pictoris MG members (blue dots) indicates the older age of these  cluster with respect to Lupus members (purple dots). Our three remaining candidates, which we acknowledge as Lupus members, have consistent $EW_{\ion{Li}{i}}$ with members of the association. Taking into consideration the ambiguities involved in analyzing the spectrum of \fiveofive \hspace{0.02cm} (see Sect. \ref{fiveofive}), we only report a three-sigma upper limit for  $EW_{\ion{Li}{i}}$ of this object.

\begin{table*}[!ht]
        \centering
        \caption{The wide-companion membership status checklist for our targets.}
        \resizebox{\textwidth}{!}{\begin{tabular}{lccccccr} 
                        \midrule
                Name & Consistent kinematic properties & Age & Active & Contains \ion{Li}{i} & conclusion \\
                & with the stellar system & & & & \\
                 & (yes/no) & (MS/PMS) & (yes/no) & (yes/no) & \\
                \midrule
                \threetn \hspace{0.02cm} & no & MS & no & no & field object \\
            \foreifor \hspace{0.02cm} & yes & PMS & yes & yes & UCL member + \\
            & & & & & HIP 74865 probable wide companion \\
                \fiveothree \hspace{0.02cm} & no & PMS & yes & yes & LCC new member \\
                \fiveofive \hspace{0.02cm} & no & PMS & yes & ambiguous & ambiguous \\
                \gqc \hspace{0.02cm} & yes & PMS & yes & yes & Lupus I new member +  \\
                & & & & & GQ Lup probable wide companion \\                     
                \midrule
        \end{tabular}}
\label{status}
\end{table*}

\begin{figure}[!ht]
\centering
\includegraphics[width=9cm]{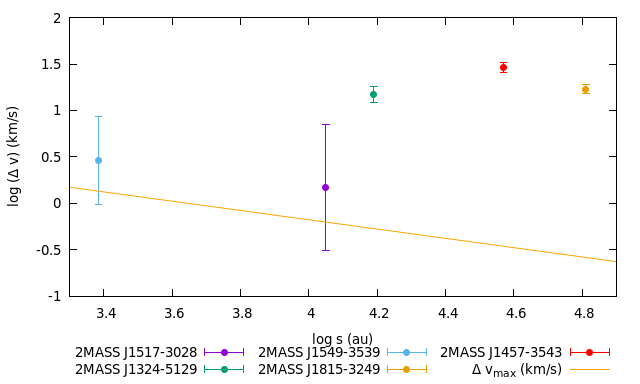}
\caption{Log--log plot of total velocity difference $\Delta v$ (km/s) vs. projected separation $s$(au) for the five plausible wide companions studied in this work. $\Delta v_{max}$ (km/s) (orange line) indicates the maximum total velocity difference that bound binaries with a total mass equal to 10 $M_{\odot}$ in circular orbits can possess; see \citet{ramirez19} for details.}
\label{fig:binding}
\end{figure}

As a final test, we also measured the total velocity difference ($\Delta v$) between our targets and their corresponding central stars. This quantity can be compared with $\Delta v_{max} (s)$, that is, the maximum total velocity difference as a function of projected separation between the two binary components in astronomical units as suggested by \citet{andrews17} for binaries of a total mass of 10 $M_{\odot}$. This comparison provides clues as to whether the possible wide companions studied in this work are truly gravitationally bound systems. As expected, we obtained large error bars associated with $\Delta v$ of \foreifor \hspace{0.02cm} and \gqc \hspace{0.02cm} (Fig. \ref{fig:binding}). The resultant values considering their large error bars are consistent with $\Delta v_{max}$ (orange line in Fig. \ref{fig:binding}), but drawing any firm conclusions regarding the physical association of the two objects is not possible because of these large error bars. This is partly due to large errors associated with the RV found for these objects and the multiplicity of the stellar systems. Thus, we still cannot conclude whether or not these two targets are gravitationally bound to their associated central stars. For \fiveothree, on the other hand, we can rule out the possibility that this target is bound to HIP 65426 as its $\Delta v$ significantly exceeds its $\Delta v_{max}$. As discussed by \citet{ramirez19}, this statement is true if there is no other effect influencing the kinematic properties of the objects involved, such as an unseen companion to any of the components or unknown systematics in \textit{Gaia DR2}.

\textbf{\foreifor \hspace{0.02cm}}:  According to our analysis, this target is a highly probable wide companion of HIP 74865 with an estimated age of 7-8 Myr --younger than the UCL sub-association (15$\pm$3 Myr) in which the object is located. This age discrepancy has been addressed previously \citep{feiden16,asensio19} as theoretical isochrones failing to reproduce the correct age of K and M stars by a factor of two (due to neglecting magnetic fields in the models). However, this issue has only been flagged up for \foreifor, and not for \fiveothree \hspace{0.02cm} or \fiveofive, \hspace{0.02cm} although these latter two are also late-type M-dwarfs, albeit later in type in comparison with \foreifor. According to \citet{baraffe15} models, the estimated ages for \fiveothree \hspace{0.02cm} and \fiveofive \hspace{0.02cm} meet our expectations and are highly consistent with the age of the stellar sub-associations they are located in. We noticed that, based on the EW of Balmer emission lines and infrared \ion{ca}{ii} triplet lines, \foreifor \hspace{0.02cm} is magnetically more active than \fiveothree \hspace{0.02cm} and \fiveofive \hspace{0.02cm}. Therefore, it might be true that the acquired age for \foreifor \hspace{0.02cm} based on \citet{baraffe15} models (that neglect the effect of magnetic fields) is younger than its real value by a factor of two. Assuming the whole stellar system as a triple system, HIP 74865 hosts a brown dwarf companion at a small separation with an M-dwarf star at a larger separation. As an already confirmed PMS member of UCL, our age estimation and determined stellar parameters reconfirm the membership of this  target. According to BANYAN $\Sigma$ \citep[][]{gagne18}, with the RV calculated for this object in the present paper, its membership probability is higher than 98\% .

\textbf{\fiveothree \hspace{0.02cm}}: Despite its similar kinematic properties to HIP 65426, their corresponding total velocity difference rules out the possibility that \fiveothree \hspace{0.02cm} is bound to this stellar system. Therefore, we conclude that \fiveothree \hspace{0.02cm} is a typical young (16$\pm$2.2 Myr) and active M-dwarf and a highly probable ---approximately 94\% according to BANYAN $\Sigma$ \citep[][]{gagne18}--- member of LCC (16$\pm$2 Myr) based on its kinematic properties and estimated age. 

\textbf{\fiveofive \hspace{0.02cm}}: We were not able to decipher whether the spectrum belongs to a single object or an unresolved binary, and therefore cannot form any firm  conclusions regarding the membership and acquired physical parameters of this object. 

\textbf{\gqc \hspace{0.02cm}}: This target is a highly probable wide companion of GQ Lupi, possibly making the stellar system a triple system consisting of a brown dwarf at a close separation (GQ Lupi b) and an M-dwarf at a larger distance. GQ Lupi b is a strong accretor with a surrounding disk, similarly to \gqc \hspace{0.02cm}. According to BANYAN $\Sigma$, the probability of \gqc \hspace{0.02cm} being a member of UCL in the Sco-Cen association is above 97\%.

Considering
the findings of the present  paper, we therefore update the status of the central stars of these systems  as follows:

\textbf{V4046 Sgr} : As its wide companion candidate, \threetn, turned out to be an unrelated field object, this stellar system remains a triple system, consisting of a very close binary with similar components (V4046 Sgr itself) and a wide companion (GSC 7396-00759). The different evolutionary phases of the disks surrounding the two objects are highly interesting and merit further investigation.

\textbf{HIP 74865 and GQ Lupi}: These two systems have similar architectures, with a BD orbiting the most massive star in the system, plus a low-mass star at a very wide separation. Besides the dynamical environment, the additional component (especially for the early type primary HIP 74865) might allow us to refine the age of the system. It is noteworthy that GQ Lup C is different in this regard: because of its strong absorption lines and edge-on disk, it is not easy to use the information offered by this target for determining the age of the system. However, in both cases there is the caveat that we are not sure if the proposed wide companions are gravitationally bound to their associated central stars, although close separations and similar kinematic properties strongly favor a common origin.

\textbf{HIP 65426}: We rule out the physical association of this star with \fiveothree. Hence, the sole companion of HIP 65426 is the planet it hosts. \citet{marleau19} modeled the formation of HIP 65426 b in the core-accretion framework and
identified the most probable path as an outward scattering after a phase of dynamical instability. The lack of wide stellar companions from \textit{Gaia} supports an origin internal to the planetary system based on this (hypothetical) framework, although a triggering by a passing star within the native association cannot be ruled out.

\textbf{HIP 73145}: This star is single as we ruled out its physical association with our candidate wide companion \fiveofive. We were hoping that finding a companion gravitationally bound to this star could provide us with hints on the formation of the multi-belt architecture of its disk. The lack of stellar and massive brown dwarf companions over the
full range of separations suggests that  these features in the disk
are not linked to external objects.

The two triple systems presented in this paper have a similar configuration, both hosting a sub-stellar object in a close orbit around the central star and a low-mass stellar companion at a larger distance. Regarding GQ Lupi, the central and companion stars are all surrounded by disks. On the contrary, for HIP 74865, no IR excess or signs of the presence of a disk around the central or companion stars are detected. As opposed to these configurations, there are other triple system architectures such as HD 284149 ABb \citep[][]{bonavita17} in which the substellar companion (HD 284149 b) has a larger separation from the central star compared to the low-mass stellar companion (HD 284149 B). There is no hint of a circumstellar disk around the central star. Various triple-system configurations of this type suggest different formation mechanisms are at play. As few such triple systems have been discovered, we cannot draw robust conclusions regarding the different formation mechanisms able to produce such configurations. However, it might be interesting to explore the currently identified stellar systems at larger distances to find out whether they possess PMS ultra-wide companions, hinting at larger multiple systems. Such studies would be valuable because multiple systems are three times more likely to have a distant companion within
10 kAU compared to single stars. On the other hand, ultra-wide companions are biased towards high multiplicity at shorter separations \citep[][]{joncour17}. If genuine, it is plausible that the aforementioned discovered systems also follow a cascade fragmentation scenario of the natal molecular core as suggested by \citet{joncour17}.

\begin{acknowledgements}
We are grateful to Dr. Julio Chanam\'e, the referee of this paper, whose comments significantly improved this work. We warmly thank G. Catanzaro for helping us in deriving the RV of HIP 73145, and S. Petrus for sharing his results on the HIP 65426 system before the publication. We thank C. F. Manara for his help during the preparation phase of the X-Shooter observations and for his useful comments and suggestions.  

This work has made use of data from the European Space Agency (ESA) mission
\textit{Gaia} (https://www.cosmos.esa.int/gaia), processed by the \textit{Gaia}
Data Processing and Analysis Consortium (DPAC,
https://www.cosmos.esa.int/web/gaia/dpac/consortium). Funding for the DPAC
has been provided by national institutions, in particular the institutions
participating in the \textit{Gaia} Multilateral Agreement.
This research has made use of the SIMBAD database and Vizier services, operated at CDS, Strasbourg, France.
This research has made use of the services of the ESO Science Archive Facility.

We acknowledge financial support from the ASI-INAF agreement n.2018-16-HH.0. JMA acknowledges financial support from INAF thourgh the project PRIN-INAF-MAIN-STREAM 2017
"Protoplanetary disks seen through the eyes of new-generation instruments".

\end{acknowledgements}

\end{document}